# Relationship between Characteristic Lengths and Effective Saffman Length in Colloidal Monolayers near a Water-Oil Interface

Na Li[1] , Wei Zhang[2], and Wei Chen[1] *

1 State Key Laboratory of Surface Physics and Department of Physics, Fudan University, Shanghai 200438, China

2 School of Physical Science and Technology, China University of Mining and Technology, Xuzhou 221116, China

Corresponding Author

*E-mail: phchenwei@fudan.edu.cn.

These authors contributed equally: Na Li, Wei Zhang.

**ABSTRACT:** The hydrodynamic interactions (HIs) in colloidal monolayers are strongly influenced by the boundary conditions and can be directly described in terms of the cross-correlated diffusion of the colloid particles. In this work, we experimentally measured the cross-correlated diffusion in colloidal monolayers near a water-oil interface. The characteristic lengths of the system were obtained by introducing an effective Saffman length. The characteristic lengths of a particle monolayer near a water-oil interface were found to be anisotropic in the longitudinal and transverse directions. From these characteristic lengths, the master curves of cross-correlated diffusion are obtained, which universally describe the HIs near a liquid-liquid interface.

**INTRODUCTION**

Confined colloidal systems are ubiquitous and important in many real scenarios. Examples of such systems include porous media[1-3], microfluidic devices[4-10], proteins in the cytoplasm[2] and cell membranes[11-13]. The dynamics of a confined colloidal suspension depend on the boundary conditions of the system[14-20] and are more complicated than those of an unbounded free 3D bulk liquid. Generally, the strength of the hydrodynamic interactions (HIs) between two tracked particles is a function of their separation distance $r$ in the liquid[3, 21]. When a membrane is suspended in a liquid, the HIs in the membrane exhibit a transition from 2D to 3D in nature with an increasing distance $r$ [21-23]. In a quasi-2D system, such as a cell membrane, a liquid-liquid interface, or a liquid film, a characteristic length can be defined to describe the distance at which the crossover from 2D membrane dynamics to 3D bulk dynamics occurs in the liquid[23-25]. When the distance $r$ is much smaller than the characteristic length, both the mass and momentum of the membrane are conserved, as in a 2D membrane[23, 26-27], which results in a decay of the HIs with the distance $r$ as $\ln(1/r)$ due to the contributions of 2D shear stress. When the distance $r$ is greater than the characteristic length, momentum is transmitted into the surrounding fluid[12, 24], which causes the HIs to decay with the distance $r$ as $1/r$ due to the contributions of 3D shear stress.

The characteristic lengths of a system depend on the boundary conditions of that system. In the Saffman model[25, 28], the characteristic length of a thin sheet of viscous fluid of viscosity $\eta_s$ suspended in an unbounded bulk liquid of viscosity $\eta_B$ is defined as $\lambda_s = \eta_s/\eta_B$, which is called the Saffman length. The characteristic length of a semi-infinite system, such as a viscous film at a water-air interface, is also taken to be equal to the Saffman length $\lambda_s = \eta_s/\eta_B$[21]. However, if the viscous film at the water-air interface is replaced by a colloidal monolayer, the characteristic length of this particle monolayer will depend on both the particle size $a$ and the Saffman length $\lambda_s$ of the monolayer in the following form: $a(\lambda_s/a)^{3/2}$[29]. When the colloidal monolayer is detached from the interface and placed in bulk water close to the water-air interface, the characteristic lengths in the two orthogonal directions are anisotropic; they are expressed as $\lambda_s$ and $a(\lambda_s/a)^{2/3}$. However, the characteristic lengths of a particle monolayer near a water-oil interface have not been closely studied until now. We wonder whether characteristic lengths of colloidal monolayers near a water-oil interface can be achieved by the method developed in colloidal monolayers on a water-air interface [29], and what different features will be exposed between those two systems. When a colloidal monolayer is located near a water-oil interface, momentum diffuses from the colloidal monolayer into the bulk water and bulk oil on either side. In such a scenario, this system can be regarded as similar to a 3D bulk system. However, the HIs in the monolayer will be influenced by the interface, such as that in monolayers near a water-air interface[30]. Thus, it is interesting to consider which kind of HI behavior can be expected in a colloidal monolayer near a water-oil interface.

In this paper, we report an investigation of the HIs in colloidal monolayers near a water-oil interface that was conducted by experimentally measuring the cross-correlated diffusion of the colloid particles. We find that an effective Saffman length $\lambda_s^*$ should be introduced in such a system. Using $\lambda_s^*$, the characteristic lengths of the colloidal monolayer can be defined, and they are found to be different in the two orthogonal directions. In the longitudinal direction, the characteristic length is equal to $\lambda_s^*$ itself, while in the transverse direction, the characteristic length exhibits a power-law relationship with $\lambda_s^*$. From these characteristic lengths, the friction coefficients of the colloidal monolayers can be estimated, and the results are consistent with those obtained from single-particle measurements. Using the characteristic lengths as scaling factors, universal curves describing the cross-correlated diffusion in a water-oil system are obtained. These universal curves are independent of the separation between the colloidal monolayer and the interface, unlike in the case of a water-air system. Such universal curves can help us to further understand the dynamics of fluids near a soft boundary.

The experimental setup is shown in Fig. 1(a). A stainless steel sample cell with two layered chambers was constructed, similar to that described in Ref. [31]. The chamber for water was 8 mm in width and 0.8 mm in depth. The chamber for oil was 15 mm in width and 1 mm in depth. An aqueous colloidal solution was introduced into the chamber for water such that it filled the chamber. Once the water-air interface was level, decalin was used to fill the oil chamber. The layers of liquid were sandwiched between two coverslips. Finally, the prepared sample was placed upside down and allowed to remain undisturbed for at least two hours. The particle solution was pinned by the edge of the stainless steel hole and remained on top of the layer of decalin. The particles sank under gravity and formed a monolayer close to the water-oil interface. The repulsive interaction of the image charges balanced the action of gravity on the particles, preventing the particles from contacting the water-oil interface. Samples of three kinds of colloid particles were used, namely, silica spheres with radii of $a = 0.6$ μm (S1) and $a = 1.0$ μm (S2) and polystyrene (PS) latex spheres with a radius of $a = 1.0$ μm (S3), as presented in Table 1. Silica spheres with $SiO^-$ groups on their surfaces were purchased from Bangs, and PS spheres with sulfate groups on their surfaces were purchased from Invitrogen. The colloidal samples were cleaned seven times by centrifugation to eliminate any possible surfactant and then suspended in deionized water with a conductivity of 18.2 MΩ·cm. The decalin used was a mixture of cis and trans with a density of 0.896 g/$cm^3$ and a viscosity of 2.36 cp at a temperature of $28^oC$, which was purchased from Sigma-Aldrich. An inverted Olympus IX71 microscope with a 40x objective and a digital camera (Prosilica GE1050, 1024*1024 pixels, 14 fps) were used to record the positions of the particles. The spatial resolution was 0.10 μm/pix. Each image sequence consisted of 500 consecutive frames, and each image had dimensions of $176 \times 176$ μm$^2$. A typical image obtained by the microscope is shown in Fig. 1(b). Under the frame-collecting rate of 14 fps, particles diffuse a short distance, which is much less than the average separation between the neighbor particles during time intervals between two adjacent frames. The closest particles in adjacent frames were identified as a same particle. Using homemade software (IDL), we obtained the trajectories of the particles by above strategy. From these particle trajectories, the single-particle diffusion and the correlated diffusion between particles were calculated to study the HIs in the colloidal monolayer.

First, from the particle trajectories, the particle displacement over a lag time $\tau$ was calculated as $\Delta \vec{s}_i(\tau) = \vec{s}_i(t+\tau) - \vec{s}_i(t)$. The single-particle self-diffusion coefficient $D_s(n)$ was calculated from the mean square displacement using the relation $\langle \Delta \vec{s}_i^2(\tau) \rangle = 4D_s(n)\tau$. As shown in Fig. 1(c), $D_s(n)$ scaled by $D_0$ is a function of the particle area fraction $n$, where $D_0$ is calculated according to the Einstein-Stokes relation: $D_0 = K_B T / 6\pi a \eta$. The measurements of this scaled diffusion coefficient $D_s(n)/D_0$ were fit to the following second-order polynomial[31-32]: $D_s(n)/D_0 = \alpha(1 - \beta \cdot n - \gamma \cdot n^2)$. Here, the parameter $\alpha$ is related to the local viscosity experienced by the particles in the dilute limit, and the parameter $\beta$ is related to the two-body interactions between the particles. When $n \to 0$, the diffusion coefficient $D_s(0)$ can be obtained from the formula $D_s(0)/D_0 = \alpha$; this coefficient is the self-diffusion coefficient of a particle in the monolayer in the dilute limit. From $D_s(0)/D_0$, the separation $z$ from the center of the colloidal monolayer to the water-oil interface was calculated according to Eq. (1)[18, 33]:

$$\frac{D_s(0)}{D_0} = 1 + \frac{3}{16}\left(\frac{2\eta_w - 3\eta_o}{\eta_w + \eta_o}\right)\left(\frac{a}{z}\right) \quad (1)$$

where $\eta_w$ is the viscosity of the water and $\eta_o$ is the viscosity of the decalin. The fitted

TABLE 1. Parameters $\alpha$, $\beta$ and $\gamma$ and the separation $z$ for the three samples

| Samples ($a$ (μm)) | $\alpha$ | $\beta$ | $\gamma$ | $z$ (μm) | $z/a$ |
|---|---|---|---|---|---|
| S1 Silica $a = 0.6$ | 0.74 | 0.55 | 2.62 | 0.73 | 1.22 |
| S2 Silica $a = 1.0$ | 0.78 | 0.80 | 0.76 | 1.46 | 1.46 |
| S3 PS $a = 1.0$ | 0.86 | 0.58 | 1.39 | 2.29 | 2.29 |

values of the parameters and the calculated values $z$ are shown in Table 1.

From the displacement $\Delta \vec{s}(\tau)$, the cross-correlated diffusion coefficients $D_{\parallel,\perp}(r)$ were calculated in the directions longitudinal and transverse to the line connecting the centers of two particles ($i$ and $j$) according to Eq. (2). In Eq. (2), $\Delta s_\parallel$ and $\Delta s_\perp$ represent the components of the displacement $\Delta \vec{s}(\tau)$ in the longitudinal and transverse directions, respectively.

$$D_{\parallel,\perp}(r) = \frac{\langle \Delta s^i_{\parallel,\perp}(t,\tau)\Delta s^j_{\parallel,\perp}(t,\tau)\delta(r-R^{i,j}(t)) \rangle_{i \neq j}}{2\tau} \qquad (2)$$

Figure 2 shows the cross-correlated diffusion coefficients $D_{\parallel,\perp}(r)$ scaled by the self-diffusion coefficient $D_s(n)$ as functions of the distance $r$, which is scaled by the particle diameter $2a$, for sample S1. The area fraction $n$ for sample S1 ranges from 0.007 to 0.20. From Fig. 2, it can be seen that the scaled cross-correlated diffusion coefficients $D_{\parallel,\perp}(r/2a)/D_s(n)$ are invariant with the concentration $n$. However, the cross-correlated diffusion should be strongly influenced by the viscosity of the colloidal monolayer, which directly depends on the concentration $n$. The effects of the concentration $n$ on the cross-correlated diffusion are hidden in the plots of $D_{\parallel,\perp}(r/2a)/D_s(n)$ for this sample. This finding suggests that the single-diffusion coefficient $D_s(n)$ and the particle diameter $2a$ are not good parameters with which to analyze the cross-correlated diffusion. The scaled cross-correlated diffusion coefficients $D_{\parallel,\perp}(r/2a)/D_s(n)$ for the other two samples, S2 and S3, are shown in Fig. S1 and Fig. S2 in the Supplementary Material.

The self-diffusion coefficient for a single particle is written as $D_s(n) = K_B T/\kappa \eta a$, where $k$ is a dimensionless constant. $\kappa \eta a$ represents the friction experienced by one particle. For a particle in a colloidal monolayer, normally such a friction include two terms, $\Gamma^{(0)}$ and $\Gamma^{(1)}$, one related to the contribution from the surrounding liquid and the other to the contribution from the colloidal monolayer. According to Ref.29-30, the self-diffusion coefficient $D_s(n)$ can be divided into two effective coefficients: $D_s(0) \equiv k_B T/\Gamma^{(0)}$ and $D'_s(n) \equiv k_B T/\Gamma^{(1)}$, where $\Gamma^{(0)} = \kappa^{(0)} \eta_b a$ and $\Gamma^{(1)} = \kappa^{(1)} \eta_s$. $\kappa^{(0)}$ and $\kappa^{(1)}$ are two dimensionless coefficients. The term $\Gamma^{(0)}$ is the friction coefficient of the colloidal monolayer in the dilute limit ($n = 0$), which reflects the influence of the local environment of the liquid. The other term, $\Gamma^{(1)}$, is the friction coefficient of the colloidal monolayer at concentrations of $n > 0$, which is influenced by the many-body effect of the particles. The effective diffusion $D_s(0)$ can be obtained by means of a fit parameter $\alpha$ as follows: $D_s(0) = \alpha \cdot D_0$. $D'_s(n)$ can be calculated from the equation $1/D_s(n) = 1/D_s(0) + 1/D'_s(n)$ [29-30]. The effective diffusion $D'_s(n)$ depends on the concentration $n$. As shown in Fig. 3, $D'_s(n)$ exhibits an inverse relation with $n$, as expressed by $D'_s/D_s(n) = 1 + 3/2n$, for each of the three samples. As shown in the inset of Fig.3, such a relationship is a general description for all samples when $n<0.1$. The value $3/2$ in above equation could be estimated by the inverse value of averaged $\beta$ in Table I.

Because $\Gamma^{(0)}$ represents the friction experienced by a single particle in dilute limit and only $\Gamma^{(1)}$ represents the friction contributed by the interaction between the particles in the monolayer[29]. Considering that $D_{\parallel,\perp}$ reflects the strength of HIs between the particles, the change of $D_{\parallel,\perp}$ induced by $n$ is contributed only by the change of $\Gamma^{(1)}$, and has nothing to do with $\Gamma^{(0)}$. After all, $\Gamma^{(1)}$ is a function of $n$, while $\Gamma^{(0)}$ is a constant and independent of $n$. Thus, the effective diffusion coefficient $D'_s(n)$, which only relates to the friction coefficient $\Gamma^{(1)}$, is a more suitable scaling factor than $D_s(n)$ is.

The plots of $D_{\parallel,\perp}/D_s$ in Fig. 2 should thus be replaced with plots of $\widetilde{D}_{\parallel,\perp} = D_{\parallel,\perp}/D'_s$. In the insets of Fig. 4(a, b), the plots of $\widetilde{D}_{\parallel,\perp}(r/2a)$ increase with increasing $\Gamma^{(1)}$, showing an influence of the concentration $n$, as expected. Moreover, new characteristic lengths should be determined to normalize the distance $r$ to obtain a universal function of cross-correlated diffusion. As shown in Fig. 4(a, b), all of the scaled cross-correlated diffusion coefficients $\widetilde{D}_{\parallel,\perp}$ collapse to single master curves when the distance $r$ is scaled by an adjustable parameter $R_{\parallel,\perp}$ to obtain $r'_{\parallel,\perp} = r/R_{\parallel,\perp}$. The scaled cross-correlated diffusion coefficients $\widetilde{D}_{\parallel,\perp}(r'_{\parallel,\perp})$ for large values of $\Gamma^{(1)}$ fall on the upper left sides of the corresponding master curves, and for small values of $\Gamma^{(1)}$, they fall on the lower right sides. The curves indicate that the scaled correlation coefficients $\widetilde{D}_{\parallel,\perp}(r'_{\parallel,\perp})$, representing the strength of the HIs, are strong for a colloidal monolayer with a high $\Gamma^{(1)}$. For the other two samples, we obtained similar results using the same method (see Fig. S3 and Fig. S4 in the supplementary material). The scaled cross-correlated diffusion coefficients $\widetilde{D}_{\parallel,\perp}(r'_{\parallel,\perp})$ for all three samples with different separations $z/a$ collapse to single master curves, as shown in Fig. 4(c, d). The universal master curves $\widetilde{D}_{\parallel,\perp}(r'_{\parallel,\perp})$ for the different samples indicate that the HIs in a particle monolayer near a water-oil interface can be described by the same rule, even for monolayers containing particles of different sizes or lying at different separation distances from the interface.

It is noted that a bit of curvature shows at small $r$ in Fig. 2(b) and Fig. 4(b). When the distance $r$ is very small, the magnitude of $r$ is close to the size of particles, i.e. particles cannot be regarded as mass points anymore, as the theoretical derivation assumed. It may cause the curvature at small $r$.

The universal master curves can be obtained only with suitable parameters $R_{\parallel,\perp}$. As length scales, $R_{\parallel,\perp}$ can be regarded as characteristic lengths of the system. The question then becomes how these characteristic lengths $R_{\parallel,\perp}$ should be interpreted. In the Saffman model[25], the characteristic length of a membrane is $\lambda_s = \eta_s/\eta_B$, which depends on the viscosity of the membrane. Based on the concept of $\lambda_s$, we define an effective Saffman length $\lambda_s^*$ for a particle monolayer near a water-oil interface, expressed as $\lambda_s^* \equiv \Gamma^{(1)}/6\pi\eta_B$. The effective Saffman length is the ratio of the friction coefficient $\Gamma^{(1)}$ of the particles in the monolayer to the traditional friction coefficient $6\pi\eta_B$ of the particles in a bulk liquid. In the present system, the effective coefficient $\lambda_s^*$ is written as $\lambda_s^* = \Gamma^{(1)}/(6\pi \cdot (\eta_w + \eta_o)/2)$. The characteristic length $R_\parallel$ in the longitudinal direction is taken to be equal to $\lambda_s^*$, as shown in Eq. (3):

$$R_\parallel = \lambda_s^* \qquad (3)$$

For given values of $R_\parallel$, corresponding friction coefficient values, denoted by $\Gamma_\parallel^{(2)}$, can be obtained from the definition of $\lambda_s^*$. The superscript and subscript of $\Gamma_\parallel^{(2)}$ suggest that this friction coefficient can be estimated from two-particle diffusion measurements in the longitudinal direction. Meanwhile, $\Gamma^{(1)}$ can also be obtained from single-particle diffusion measurements, in accordance with the formulas $D_s'(n) = k_B T/\Gamma^{(1)}$ and $1/D_s(n) = 1/D_s(0) + 1/D_s'(n)$. As shown in Fig. 5(a), the $\Gamma_\parallel^{(2)}$ values obtained from $R_\parallel$ agree with the $\Gamma^{(1)}$ values obtained from $D_s'(n)$, which justifies the rationality of the relationship $R_\parallel = \lambda_s^*$. However, in the transverse direction, the equivalent of Eq. (3), $R_\perp = \lambda_s^*$, does not work as the characteristic length. If $R_\perp = \lambda_s^*$ is assumed, then the friction coefficient $\Gamma_\perp^{(2)}$ obtained from $R_\perp$ is not a linear function of $\Gamma^{(1)}$. In fact, $\Gamma_\perp^{(2)}$ will exhibit a power-law relationship with $\Gamma^{(1)}$, expressed as $\Gamma_\perp^{(2)} \sim (\Gamma^{(1)})^{2/3}$. Such a wrong dependence of $\Gamma_\perp^{(2)}$ on $\Gamma^{(1)}$ given by the assumption ($R_\perp = \lambda_s^*$) is shown in Fig. 5(b), as a comparison of the correct dependence in Fig. 5(a). As suggested by this power-law relationship, the transverse characteristic length $R_\perp$ should instead be written as

$$R_\perp = a\left(\frac{\lambda_s^*}{a}\right)^{2/3} \qquad (4)$$

for $\Gamma_\perp^{(2)} = \Gamma^{(1)}$ to be satisfied. The plot of $\Gamma_\perp^{(2)}$ vs. $\Gamma^{(1)}$ is shown in Fig. 5(a). Both $\Gamma_\parallel^{(2)}$ and $\Gamma_\perp^{(2)}$ are friction coefficients obtained for the same colloidal monolayer. Unsurprisingly, $\Gamma_\parallel^{(2)}$ and $\Gamma_\perp^{(2)}$ should be equal to each other, as shown in the inset of Fig. 5(a). Thus, we can define the friction coefficient obtained from cross-correlated diffusion as $\Gamma^{(2)}$, where $\Gamma^{(2)} = \Gamma_\parallel^{(2)} = \Gamma_\perp^{(2)}$. The characteristic lengths $R_\parallel$ and $R_\perp$ follow

Eq. (3) and Eq. (4), respectively, indicating that the HIs are anisotropic in the two orthogonal directions. By combining Eq. (3) and Eq. (4), the relationship between $R_\parallel$ and $R_\perp$ can be found to be $R_\parallel/a = (R_\perp/a)^{3/2}$, as shown in Fig. 5(c).

In Fig. 2(a,b), $D_\parallel/D_s(n)$ for sample S1 decays with the distance $r$ as $\sim 1/r$ in the limit of large $r$, while $D_\perp/D_s(n)$ decays as $\sim 1/r^2$. These behaviors are the result of the anisotropic HIs in a colloidal monolayer near a liquid-liquid interface. In the longitudinal direction, the HIs respond to a 3D shear stress that is similar to that in a 3D bulk liquid, and in the transverse direction, the HIs respond to a 2D compressive stress[12, 24]. The anisotropy of the characteristic lengths $R_\parallel$ and $R_\perp$ of a colloidal monolayer near a water-oil interface is also similar to that near a water-air interface [30]. This anisotropic behavior can be explained by analogy to the lubrication force between two solid walls[30, 34].

In the system of a water-oil interface, the scaled correlated diffusion coefficients $\widetilde{D}_{\parallel,\perp}(r'_{\parallel,\perp})$ are independent on the separation $z$, as seen in Fig. 4(c, d), while $\widetilde{D}_{\parallel,\perp}(r'_{\parallel,\perp})$ in the water-air system are dependent on the separation $z$ [30]. Comparing these two systems, the significant difference is that the viscosity and mass density of the air can be negligible comparing with oil. There is no clear influence of $z$ on $\widetilde{D}_{\parallel,\perp}$ curves in fig. 4. HIs contributed by the oil overwhelms small change of HIs resulting from the different locations of colloidal monolayers. In the colloidal monolayer near the water-air interface, the momentum (stress flux) of the monolayer diffuse only into one side of the interface, i.e. the region of water. The HIs near the water-air interface is sensitive to the thickness of the water film between the monolayer and the interface. In the water-oil system, the momentum (stress flux) of the monolayer diffuse into both sides of the interface, i.e. the regions of water and oil. Comparing with the influence of the bulk of water and oil, the influence of the thin water film between the monolayer and the water-oil interface can be negligible, since z is only of the order of µm. The diffusion of the momentum (stress flux) into the whole space causes that $\widetilde{D}_{\parallel,\perp}(r'_{\parallel,\perp})$ in the water-oil system is insensitive to the separation $z$.

The high viscosity of oil may also cause the curves collapse Fig. 2 (Sample S1) and Fig. S2 (Sample S3), since such collapse never found in monolayers near a water-air interface[30]. We think the difference of $D_{\parallel,\perp}$ induced by the change of particle concentrations lie already in $D_s(n)$, but only in the part of $D'_s(n)$. $D_s(0)$ is just a constant and independence on $n$. It is the reason why the rescaling method in Fig. 4 works so well for all the situations.

In this work, we measured the cross-correlated diffusion in particle monolayers near a water-oil interface. An effective Saffman length of $\lambda_s^* \equiv \kappa^{(1)} \eta_s / 6\pi \eta_B$ was defined for such a colloidal monolayer. On the basis of this effective Saffman length, anisotropic characteristic lengths $R_\parallel$ and $R_\perp$ were introduced in the longitudinal and transverse directions, respectively. In the longitudinal direction, the characteristic length $R_\parallel$ is equal to $\lambda_s^*$, which reflects the continuous nature of the system. In the transverse direction, the characteristic length $R_\perp$ is a function of $\lambda_s^*$ and the particle radius $a$, with the form $a(\lambda_s^*/a)^{2/3}$, which reflects the discrete nature of the system. The anisotropy of the characteristic lengths reflects the fundamental differences in the dynamics of the colloidal monolayer in the two directions. With these two characteristic lengths, the measured curves of the cross-correlated diffusion for three samples at different concentrations collapse to universal master curves. These universal master curves describe the HIs in a particle monolayer near a soft interface. This experimental research provides a general picture of the dynamics of a liquid near an interface.

**Acknowledgment**

This research is supported by the National Natural Science Foundation of China (Grant No. 11474054). The author Wei Zhang also thanks for the support of National Natural Science Foundation of China (11774417 and 11604381), and the Natural Science Foundation of Jiangsu Province (Grant No. BK20160238).

**Author Contributions**

Wei Chen designed the experiments and paper writing. Na Li performed the experiments and paper writing. Wei Zhang contributed to perform the experiments and the preparation of the manuscript.

**Notes**

The authors declare no competing financial interest.

Figures

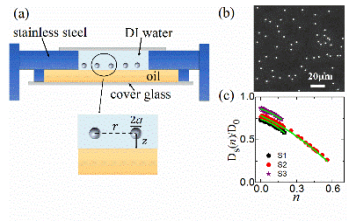

FIG. 1. (a) Schematic illustration of the experimental setup. The distance between particles is denoted by $r$, the separation from the center of the particle monolayer to the water-oil interface is denoted by $z$, and the parameter $a$ is the particle radius. (b) Optical microscopic image of silica particles of radius $a = 1.0$ μm suspended near a water-decalin interface at an area fraction of $n = 0.01$. (c) Normalized self-diffusion coefficient $D_s(n)/D_0$ as a function of the area fraction $n$ for each of the three samples: S1 (black pentagon), S2 (red dot), and S3 (purple star). The green curves are fits to the following second-order polynomial: $D_s(n)/D_0 = \alpha(1 - \beta \cdot n - \gamma \cdot n^2)$.

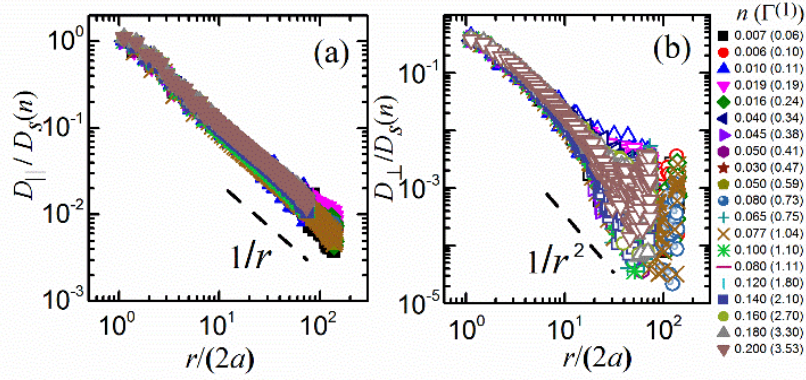

FIG. 2. Measured correlated diffusion coefficients $D_{\parallel}/D_s(n)$ (a) and $D_{\perp}/D_s(n)$ (b) as functions of $r/(2a)$ for sample S1 at different particle area fractions $n$ on a log-log plot. The concentration $n$ ranges from 0.007 to 0.20, and the friction coefficient of the particle monolayer ranges from 0.06 cP μm to 3.53 cP μm. The differently colored symbols represent different values of $n$ and different viscosity coefficients, where the black squares correspond to $n \approx 0.007$ and the maroon inverted triangles correspond to $n \approx 0.20$. The dashed lines are guides for the eye, corresponding to $\sim 1/r$ in (a) and to $\sim 1/r^2$ in (b).

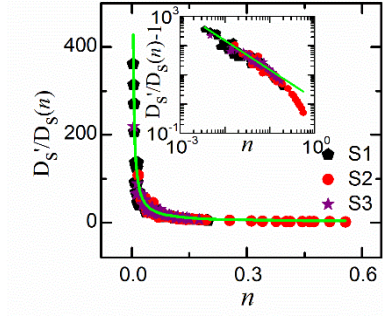

FIG. 3. The calculated $D'_s(n)$ values scaled by $D_s(n)$ as a function of the area fraction $n$ for all three samples: S1 (black pentagon), S2 (red dot), and S3 (purple star). The green line shows the fit to $D'_s(n)/D_s(n) = 1 + 3/(2n)$. A log-log replot of $D'_s(n)/D_s(n) - 1\ vs. n$ is shown in the inset, where a green line with a slope of -1 is plotted as eye guide.

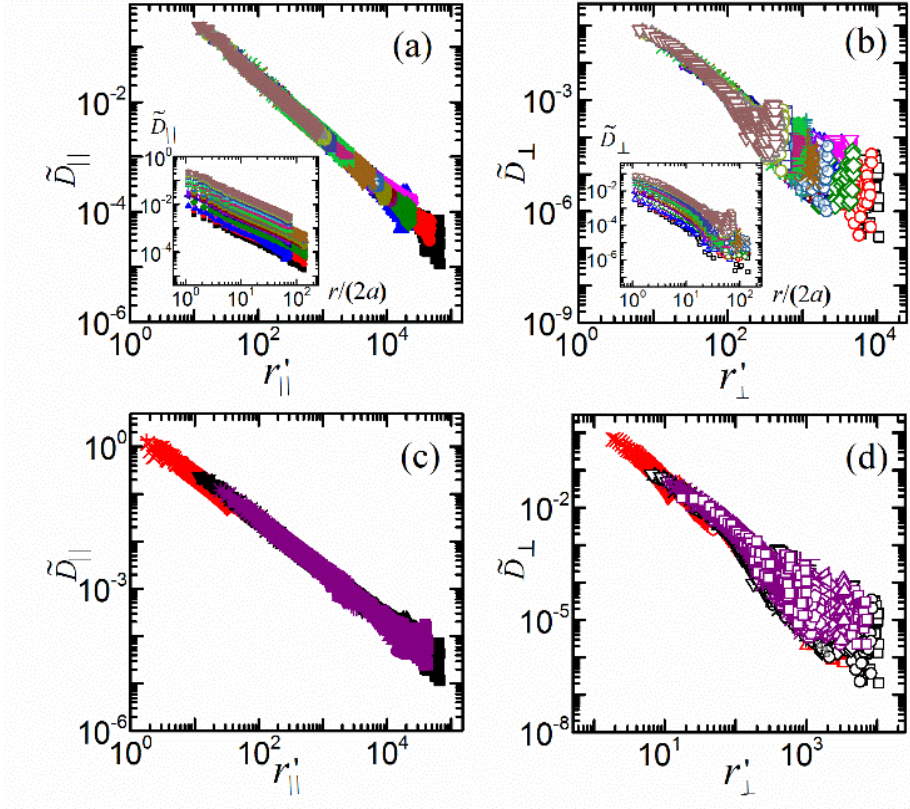

FIG. 4. (a) Scaled correlated diffusion coefficient $\widetilde{D}_\parallel$ as a function of the scaled distance $r'_\parallel$ for sample S1. (b) Scaled correlated diffusion coefficient $\widetilde{D}_\perp$ as a function of the scaled distance $r'_\perp$ for sample S1. Insets: Scaled correlated diffusion coefficients $\widetilde{D}_\parallel$ (a) and $\widetilde{D}_\perp$ (b) as functions of $r/(2a)$. The symbols used in (a) and (b) are the same as those in Fig. 2. (c) Universal master curve of $\widetilde{D}_\parallel$ as a function of $r'_\parallel$ for all three samples: S1 (black), S2 (red), and S3 (purple). (d) Universal master curve of $\widetilde{D}_\perp$ as a function of $r'_\perp$ for all three samples.

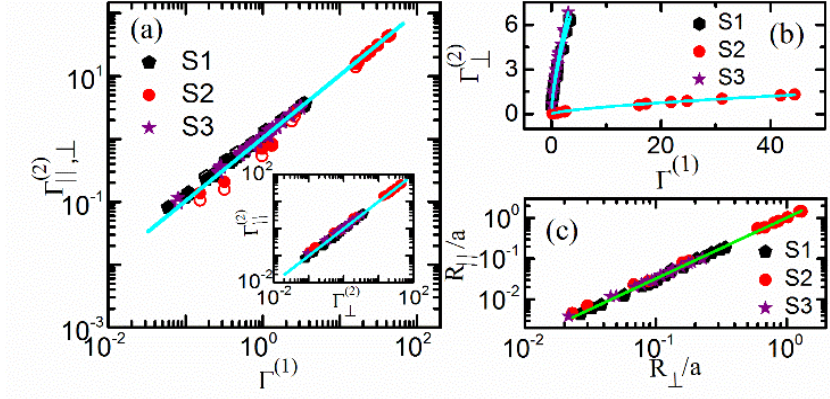

FIG. 5. (a) Comparison between the friction coefficients $\Gamma^{(1)}$ calculated from $D'_s(n)$ and the friction coefficients $\Gamma^{(2)}_{\parallel,\perp}$ of the three samples S1 (black pentagon), S2 (red dot), and S3 (purple star) on a log-log plot. The solid dots correspond to $\Gamma^{(2)}_{\parallel}$, and the open dots correspond to $\Gamma^{(2)}_{\perp}$. The cyan line is the guide for the eye, where the slope of the line is 1.0. In the inset, a comparison is shown between the friction coefficients $\Gamma^{(2)}_{\parallel}$ and $\Gamma^{(2)}_{\perp}$ on a log-log plot. The cyan line is the guide for the eye, where the slope of the line is 1.0. (b) The friction coefficient $\Gamma^{(2)}_{\perp}$ obtained from $R_{\perp} = \lambda_s$ follows a power-law relationship with the friction coefficient $\Gamma^{(1)}$ for the three samples. The cyan lines represent fits of the form $\Gamma^{(2)}_{\perp} \sim (\Gamma^{(1)})^{2/3}$. (c) Relationship between the characteristic lengths $R_{\parallel}$ and $R_{\perp}$ scaled by the particle radius $a$ for the three samples on a log-log plot. The green line is the fit to $R_{\parallel}/a = (R_{\perp}/a)^{3/2}$.

TOC graph

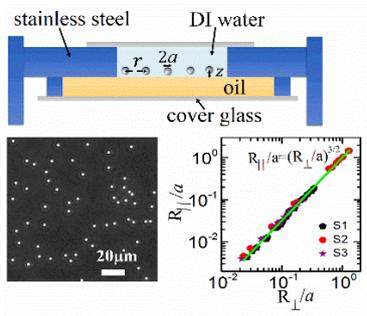

TOC graph: A colloidal monolayer near a water-oil interface. Relationship between the characteristic lengths $R_{\parallel,\perp}$ of the colloidal monolayer in the two orthogonal directions.

BRIEFS

The hydrodynamic interactions between colloidal particles are influenced by a liquid-liquid interface, which can be reflected in the behavior of correlated diffusion coefficients of particles. In the system of the colloidal monolayer near a water-oil interface, an effective Saffman length is introduced. On the basis of this effective Saffman length, anisotropic characteristic lengths $R_{\parallel}$ and $R_{\perp}$ were defined to describe the hydrodynamic interactions of the colloidal monolayer near a liquid-liquid interface